\documentstyle[12pt]{article}

\newtheorem{Theorem}{Theorem}
\newtheorem{Lemma}{Lemma}
\newtheorem{Claim}{Claim}
\newtheorem{Corollary}{Corollary}

\def\bbbn{{\rm I\!N}}
\def\bbbr{{\rm I\!R}}
\def\bbbc{{\mathchoice {\setbox0=\hbox{$\displaystyle\rm C$}\hbox{\hbox
to0pt{\kern0.4\wd0\vrule height0.9\ht0\hss}\box0}}
{\setbox0=\hbox{$\textstyle\rm C$}\hbox{\hbox
to0pt{\kern0.4\wd0\vrule height0.9\ht0\hss}\box0}}
{\setbox0=\hbox{$\scriptstyle\rm C$}\hbox{\hbox
to0pt{\kern0.4\wd0\vrule height0.9\ht0\hss}\box0}}
{\setbox0=\hbox{$\scriptscriptstyle\rm C$}\hbox{\hbox
to0pt{\kern0.4\wd0\vrule height0.9\ht0\hss}\box0}}}}

\begin{document}

\date{}

\title{\bf A better lower bound for quantum algorithms searching an ordered list}

\author{Andris Ambainis\\
       Computer Science Division\\
       University of California\\
       Berkeley, CA 94720,\\
       e-mail: {\tt ambainis@cs.berkeley.edu}\thanks{Supported 
       by Berkeley Fellowship for Graduate Studies.}}

\maketitle

\begin{abstract}
We show that any quantum algorithm searching an ordered
list of $n$ elements needs to examine at least $\frac{\log_2 n}{12}-O(1)$
of them. Classically, $\log_2 n$ queries are both necessary and
sufficient. This shows that quantum algorithms can achieve only
a constant speedup for this problem. Our result improves lower bounds of
Buhrman and de Wolf(quant-ph/9811046) and Farhi, Goldstone, Gutmann
and Sipser (quant-ph/9812057). 
\end{abstract}

\section{Introduction}

One of main results in quantum computation is Grover's 
algorithm\cite{Grover}. This quantum algorithm allows to search an unordered
list of $n$ elements by examining just $O(\sqrt{n})$ of them.
Any classical algorithm needs to examine all $n$ elements.
Grover's algorithm is very important because it can be applied
to any search problem (not just searching a list).
For example, it can be used to find a Hamilton cycle in an $n$-vertex
graph by checking only $\sqrt{n!}$ out of $n!$ possible Hamilton cycles.

After Grover's paper appeared, unordered search and related problems
received a lot of attention in quantum computation community.
It was shown that $O(\sqrt{n})$ is optimal\cite{BBBV}.
Then, Grover's algorithm has been
used as subroutine in other quantum algorithms\cite{Counting,Collision}.
Optimality proof of \cite{BBBV} has been generalized as 
well\cite{Polynomials,BW}.

Grover's algorithm works well for unordered lists but cannot be
used for searching ordered lists. An ordered list of $n$ elements
can be searched by examining just $\log_2 n$ elements classically
and there is no evident way of speeding it up by methods similar to
Grover's algorithm. 

Searching ordered lists by a quantum algorithm
was first considered by Buhrman and de Wolf\cite{BW} who proved
a $\sqrt{\log n}/\log \log n$ lower bound for quantum case.
This lower bound was improved to $\log n/2\log\log n$ by Farhi, Goldstone,
Gutmann and Sipser\cite{Sipser}. The $\log n/2\log\log n$ bound
was independently discovered (but not published) by the author of
this paper in June 1998. We improve the lower bound to $\frac{1}{12} \log n-O(1)$,
showing that only a constant speedup is possible for the
ordered search.

The best quantum algorithm for ordered search uses
$0.53\log n$ queries\cite{Sipser2}.\footnote{A different quantum algorithm 
with $c\log n$ queries for $c<1$ was claimed in \cite{Rohrig}.
However, a bug was discovered in the proof of \cite{Rohrig}
and it is not clear whether the proof can be fixed.}
Thus, a constant speedup is possible. 

The proof of our lower bound combines the method of \cite{BBBV}
(adapted to ordered case by \cite{Sipser}) with a new idea
inspired by weighted majority algorithms in the learning
theory\cite{BFP,LW}.

\section{Preliminaries}

\subsection{Quantum binary search}

In the binary search problem,
we are given $x_1\in\bbbr, \ldots, x_n\in\bbbr, y\in\bbbr$ such that
$x_1\leq x_2\leq\ldots\leq x_n$ and have to find 
the smallest $i$ such that $y\leq x_i$. 
Normally, $x_1$, $\ldots$, $x_n$ are accessed by queries.
The input to the query is $i$, the answer is $x_i$.
This is a classical problem in computer science and it
is well known that $\log_2 n$ queries are both necessary and sufficient
to solve it classically.

In this paper, we consider how many queries one needs in the quantum world.
We will prove a $\frac{1}{12}\log n-O(1)$ lower bound.
For our proof, it is enough to consider the case when
$x_1\in\{0, 1\}$, $\ldots$, $x_n\in\{0, 1\}$ and $y=1$.
Then the problem becomes following.

{\bf 0-1 valued binary search.}
Given $x_1\in\{0, 1\}$, $\ldots$, $x_n\in\{0, 1\}$ such that
$x_1\leq x_2\leq\ldots\leq x_n$, find the smallest $i$ such that $x_i=1$.

Similarly to classical world, we consider algorithms that access
the input by queries. A quantum algorithm $A$ 
can be represented as a sequence of unitary transformations 
\[ U_0\rightarrow O \rightarrow U_1\rightarrow O\rightarrow\ldots
\rightarrow U_{T-1}\rightarrow O\rightarrow U_T\]
on a state space with finitely many basis states.
$U_j$'s are arbitrary unitary transformations that do not depend
on $x_1, \ldots, x_N$ and $O$'s are queries to the input.
We use $O_k$ to denote the query transformation corresponding 
to the input $x_1=\ldots=x_k=0$, $x_{k+1}=\ldots=x_n=1$.

To define $O_k$, we represent the basis states as $|i, b, z\rangle$ where 
$i$ consists of $\lceil \log n\rceil$ bits, $b$ is one bit and 
$z$ consists of all other qubits. Then, $O_k$ maps
$|i, b, z\rangle$ to $|i, b\oplus x_i, z\rangle$.
(I.e., the first $\lceil\log n\rceil$ qubits are interpreted as an index
$i$ for an input bit $x_i$ and this input bit is XORed on 
the next qubit.) 

Running a quantum algorithm $A$ on an input
$x_1=\ldots=x_k=0$, $x_{k+1}=\ldots=x_n=1$ means applying
the transformation $U_T O_k U_{T-1} \ldots U_1 O_k U_0$
to the initial state $|0\rangle$ and measuring the first
$\lceil\log n\rceil$ bits of the final state.
The algorithm computes the binary search function if,
for any input $x_1=\ldots=x_k=0$, $x_{k+1}=\ldots=x_n=1$,
this process gives $k$ with probability at least 
3/4\footnote{One can replace 3/4 by any other 
constant in (0, 1) and our proof would still give 
a $\frac{1}{12}\log n-O(1)$ lower bound, with a slightly
different constant in the $O(1)$ term.}.

\subsection{Technical lemmas}

In this section, we state several results that we will use.
The first result is the well-known formula for the sum of 
decreasing geometric progression. If $q>1$, then
\begin{equation}
\label{geom} 
\sum_{i=0}^{\infty} \frac{1}{q^i}=\frac{1}{1-\frac{1}{q}}, 
\mbox{                            }
\sum_{i=1}^{\infty} \frac{1}{q^i}=\frac{1}{q-1} 
\end{equation}

The second result is a lemma from \cite{BV}. 
It relates the $l_2$-distance between
two superpositions and the variational distance between probability 
distributions that we obtain by observing two superpositions.
The variational distance between two probability
distributions $p(x)$ and $p'(x)$
is just the sum $\sum_{x}|p(x)-p'(x)|$.

\begin{Lemma}\footnote{Ronald de Wolf has shown that $4\epsilon$
can be improved to $2\epsilon$ in this lemma. This can be used to
improve $O(1)$ constant in our $\frac{1}{12}\log n-O(1)$ lower
bound.}
\cite{BV}
\label{BVTheorem}
Let $\psi$ and $\phi$ be superpositions
such that $\|\psi-\phi\|\leq\epsilon$.
Then the total variational distance resulting from measurements
of $\phi$ and $\psi$ is at most $4\epsilon$.
\end{Lemma}

In our case, $\psi$ and $\phi$ are final superpositions
of a quantum algorithm $A$ on two different
inputs $x_1=\ldots=x_j=0$, $x_{j+1}=\ldots=x_n=1$
and $x_1=\ldots=x_k=0$, $x_{k+1}=\ldots=x_n=1$.
For the first input, $j$ is the correct answer and the measurement
must return $j$ with probability at least $\frac{3}{4}$.
The probability that the measurement gives $k$ can be
at most $1-\frac{3}{4}=\frac{1}{4}$.
For the second input, the probability of $j$ can be at most 
$\frac{1}{4}$ and the probability of $k$ must be at least $\frac{3}{4}$.
This means that the variational distance must be
at least $2(\frac{3}{4}-\frac{1}{4})=1$.
By Lemma \ref{BVTheorem}, this is only possible if 
$\|\psi-\phi\|\geq \frac{1}{4}$.
We have shown

\begin{Lemma}
\label{distance}
If $\psi$ and $\phi$ are final superpositions of a quantum
binary search algorithm, then $\|\psi-\phi\|\geq\frac{1}{4}$.
\end{Lemma}

\section{Result}

\subsection{$\log n/2\log\log n$ lower bound}

We start with a sketch of $\log n/2\log\log n$ lower bound discovered
independently by the author of this paper and Farhi, Gutmann, Goldstone
and Sipser\cite{Sipser}. After that, we describe how to
modify this argument to obtain an $\Omega(\log n)$ lower bound.

Assume we are given a quantum algorithm $A$ for binary search that
uses less than $\log n/2 \log\log n$ queries. We construct an input 
on which $A$ works incorrectly. In the first stage, we
partition $[1, n]$ into $\log^2 n$ intervals of length $n/\log^2 n$ each.
We simulate $A$ up to the first query. 
There is an interval $[(l-1)\cdot n/\log^2 n+1, l\cdot n/\log^2 n]$  
that is queried with probability that less than or equal to $1/\log^2 n$.
We answer the first query of $A$ with $x_i=0$ for 
$i\leq l\cdot n/\log^2 n$ and $x_i=1$ for $i> l\cdot n/\log^2 n$.
Then, we split the interval $[(l-1)\cdot n/\log^2 n+1, l\cdot n/\log^2 n]$
into $\log^2 n$ parts of size $n/\log^4 n$, find the one queried with
the smallest probability, answer the second query by $x_i=0$ for $i$ 
up to this interval and $x_i=1$ for greater $i$ and so on.
We repeat the splitting until the interval is smaller than $\log^2 n$.
This means doing $\log n/\log(\log^2 n)=\log n/2\log\log n$
splittings.

Let $[(l-1)m+1, lm]$ be the final interval.
Consider two inputs $x_1=\ldots=x_{lm-1}=0$, 
$x_{lm}=\ldots=x_n=1$ and $x_1=\ldots=x_{lm}=0$, 
$x_{lm+1}=\ldots=x_n=1$.
The only value where these two inputs differ is $x_{lm}$ and,
by our construction, it is queried with a probability at most 
$1/\log^2 n$ in each of $\log n/2\log\log n$ steps.
By a hybrid argument similar to \cite{BBBV},
this implies that the final superpositions of the quantum algorithm
$A$ on these two inputs are within distance $O(1/\log\log n)$. 
Hence, the results of measuring  
final superpositions on two inputs will be
close as well (cf. Lemma \ref{BVTheorem}).

\subsection{$\log n/12$ lower bound}

To obtain an $\Omega(\log n)$ lower bound, we must split the interval into
a constant number of pieces at every step (rather than $\log^2 n$ pieces).
However, if we split the interval into a constant number of pieces,
we can only guarantee that the new interval has the probability
of being queried smaller than some constant (not smaller than $1/\log^2 n$).
Then, it may happen that $x_{lm}$ gets queried with a constant probability
in each of $c\log n$ queries, giving the total probability much higher than 1.
In this case, the quantum algorithm $A$ can easily distinguish 
two inputs that differ only in $x_{lm}$.

To avoid this, we do the splitting in a different way.
Instead of considering just the probabilities
of an interval being queried in the last step, we consider the probabilities 
of it being queried in the previous steps as well and try to decrease them all.
This is done by using a weighted sum of these probabilities.
The precise argument follows.

\begin{Theorem}
\label{main}
Let $q\in\bbbr$, $q>1$, $t\in\bbbn$, $u\in\bbbn$ and
\[ q'= \frac{1}{\sqrt{t}}+\frac{2}{q-1}\]
be such that $q (q')^u <1$.
Then, at least $\frac{\log n}{u \log t}-O(1)$
queries are necessary for the quantum binary search on $n$ elements.
\end{Theorem}
 
\noindent
{\bf Proof:}
Assume we are given a quantum algorithm $A$ doing binary search on 
$x_1, \ldots, x_n$ with less than $\frac{\log n}{u\log t}-c$
queries where the constant $c$ will be specified later.
We construct two inputs that $A$ cannot distinguish.

First, we describe an auxiliary procedure $subdivide$.
This procedure takes an interval $[(l-1)m+1, lm]$ and
returns a subinterval $[(l'-1)m'+1, l'm']$.

$subdivide(m, l, s)$:

\begin{enumerate}
\item
Let $m'=m/t$.
Split $[(l-1)m+1, lm]$ into $t$ subintervals:
$[(l-1)m+1, (l-1)m+m']$, $[(l-1)m+m'+1, (l-1)m+2m']$,
$\ldots$, $[(l-1)m+(t-1)m'+1, (l-1)m+tm']$.
\item
Simulate the first $s$ query steps (and unitary transformations 
between these steps) of $A$ on the input
$x_1=\ldots= x_{l m}=0$, $x_{lm+1}=\ldots=x_{n}=1$.
Let 
\[ |\phi_{i}\rangle= U_{i-1} O_{lm} U_{i-2}\ldots U_1 O_{lm} U_0 (|0\rangle)\] 
be the superposition before the $i^{\rm th}$ query
step, $|\psi_i\rangle$ be its part corresponding to querying $x_k$ for
$k\in[(l-1)m+1, lm]$ and $|\psi_{i, r}\rangle$ be the part of 
$|\psi_i\rangle$ corresponding to querying $x_k$ for 
$k\in[(l-1)m+(r-1)m'+1, (l-1)m+rm']$.
\item
For every $r\in\{1, \ldots, t\}$,
compute the sum 
\[ S_r=\||\psi_{s, r}\rangle\|+q \||\psi_{(s-1), r}\rangle\|+ 
q^2 \||\psi_{(s-2), r}\rangle\|
+\ldots+ q^{s-1}\||\psi_{1, r}\rangle\|.\]
\item
Take the $r$ minimizing $S_r$ and set
$l'=(l-1) k+r$. Then, $[(l'-1)m'+1, l'm']$ is equal to 
$[(l-1)m+(r-1)m'+1, (l-1)m+rm']$.
\end{enumerate}

Next, we analyze this procedure.
Let 
\[ S=\||\psi_s\rangle\|+ q \||\psi_{s-1}\rangle\|+
q^2\||\psi_{s-2}\rangle\|+\ldots+q^{s-1}\||\psi_{1, j}\rangle\|.\]
Let $\phi'_i$ and $\psi'_i$ be the counterparts for
$\phi_i$ and $\psi_i$, given the input
$x_1=\ldots= x_{l'm'}=0$, $x_{l'm'+1}=\ldots=x_{n}=1$.
We define $S'=\||\psi'_s\rangle\|+ q \||\psi'_{s-1}\rangle\|+
q^2\||\psi'_{s-2}\rangle\|+\ldots$.

\begin{Lemma}
\label{L1}
\[ S'\leq q' S .\]
\end{Lemma}

\noindent
{\bf Proof:}
We bound the difference between superpositions $|\psi'_{i}\rangle$
(used to define $S'$) and $|\psi_{i, r}\rangle$ (used to define $S_r$).
To do that, we first bound the difference between
$|\phi_i\rangle$ and $|\phi'_i\rangle$.

\begin{Claim}
\[ \||\phi_i\rangle-|\phi'_i\rangle\|\leq 
2(\||\psi_1\rangle\|+\ldots+\||\psi_{i-1}\rangle\|) .\]
\end{Claim}

\noindent
{\bf Proof:}
By induction. 
If $i=1$, then $|\phi_i\rangle=|\phi'_i\rangle$.

Next, we assume that $\||\phi_{i-1}\rangle-|\phi'_{i-1}\rangle\|\leq 
2(\||\psi_1\rangle\|+\ldots+\||\psi_{i-2}\rangle\|)$.
$\phi_i$ is the result of applying $U_i O_{lm}$ to $|\phi_{i-1}\rangle$
and $|\phi'_i\rangle$ is the result of applying $U_i O_{l'm'}$ to 
$|\phi'_{i-1}\rangle$. $U_i$ is just a unitary transformation and it
does not change distances. Hence, we have
\[ \||\phi_i\rangle-|\phi'_i\rangle\|
=\|U_i O_{lm}(|\phi_{i-1}\rangle)-U_i O_{l'm'}(|\phi'_{i-1}\rangle)\|=
\| O_{lm}(|\phi_{i-1}\rangle)-O_{l'm'}(|\phi'_{i-1}\rangle)\| \leq\] 
\[ \| O_{lm}(|\phi_{i-1}\rangle)-O_{l'm'}(|\phi_{i-1}\rangle)\| + 
\| O_{l'm'}(|\phi_{i-1}\rangle)-O_{l'm'}(|\phi'_{i-1}\rangle)\| .\]
The second part is just $\||\phi_{i-1}\rangle-|\phi'_{i-1}\rangle\|$ and
it is at most $2(\||\psi_1\rangle\|+\ldots+\||\psi_{i-2}\rangle\|)$
by inductive assumption.
To bound the first part, let 
$|\varphi_{i-1}\rangle=|\phi_{i-1}\rangle-|\psi_{i-1}\rangle$.
Then, $|\varphi_{i-1}\rangle$ is the part of superposition 
$|\phi_{i-1}\rangle$ corresponding to querying $k\notin[(l-1)m+1, lm]$.
$O_{lm}$ and $O_{l'm'}$ are the same for such $k$.
Therefore, 
$O_{lm}(|\varphi_{i-1}\rangle)=O_{l'm'}(|\varphi_{i-1}\rangle)$
and 
\[ \| O_{lm}(|\phi_{i-1}\rangle)-O_{l'm'}(|\phi_{i-1}\rangle)\| = 
\| O_{lm}(|\psi_{i-1}\rangle)-O_{l'm'}(|\psi_{i-1}\rangle)\| \leq\]
\[ \|O_{lm}(|\psi_{i-1}\rangle)\|+
\| O_{l'm'}(|\psi_{i-1}\rangle)\| = 2\||\psi_{i-1}\rangle\|. \]
$\Box$

Consider the subspace of the Hilbert space consisting of 
states that correspond to querying $k\in[(l'-1)m', l'm']$.
$|\psi_{i, r}\rangle$ and $|\psi'_i\rangle$ are projections of 
$|\phi_i\rangle$ and $|\phi'_i\rangle$ to this subspace. 
Hence, $|\psi_{i, r}\rangle-|\psi'_i\rangle$ is the projection 
of $|\phi_i\rangle-|\phi'_i\rangle$.
For any vector, the norm of its projection is at most 
the norm of the vector itself.
Therefore, we have

\begin{Claim}
\[ \||\psi_{i, r}\rangle-|\psi'_i\rangle\|
\leq 2(\||\psi_1\rangle\|+\ldots+\||\psi_{i-1}\rangle\|) .\]
\end{Claim}

This means 
\[ S'=\sum_{i=1}^s q^{s-i} \||\psi'_{i}\rangle\| 
\leq \sum_{i=1}^s q^{s-i} (\||\psi_{i, r}\rangle\|+
2(\||\psi_1\rangle\|+\ldots+\||\psi_{i-1}\rangle\|) ) =\]
\begin{equation}
\label{e3} 
\sum_{i=1}^s q^{s-i} \||\psi_{i, r}\rangle\| +
2 \sum_{i=1}^s ( q^{s-i} \sum_{j=1}^{i-1} \||\psi_j\rangle\| ) .
\end{equation}
The first term is just $S_r$. Next, we bound the second term.
\[ \sum_{i=1}^s (q^{s-i} \sum_{j=1}^{i-1} \||\psi_j\rangle\|) 
=\sum_{j=1}^s (\||\psi_j\rangle\| \sum_{i=j+1}^s q^{s-i})
< \sum_{j=1}^s (\||\psi_j\rangle\| \sum_{i=j+1}^{\infty} q^{s-i})=\]
\begin{equation}
\label{e4} 
\sum_{j=1}^s (\||\psi_j\rangle\| q^{s-j} \sum_{i=1}^{\infty} 
\frac{1}{q^i}) =
\sum_{j=1}^s \||\psi_j\rangle\| \frac{q^{s-j}}{q-1}=\frac{1}{q-1} S .
\end{equation} 
Putting (\ref{e3}), (\ref{e4}) together, we get $S'\leq S_r+\frac{2}{q-1}S$.
Next, we bound $S_r$.

\begin{Claim}
\label{C3}
\begin{equation}
\label{e5} 
S_r\leq\frac{1}{\sqrt{t}}S.
\end{equation}
\end{Claim}

\noindent
{\bf Proof:}
We have $|\psi_i\rangle=|\psi_{i, 1}\rangle+\ldots+
|\psi_{i, t}\rangle$. 
This implies
\[ \||\psi_i\rangle\|^2=
\||\psi_{i, 1}\rangle\|^2+\ldots+\||\psi_{i, t}\rangle\|^2 .\]
By a Cauchy-Schwartz inequality
\[ \||\psi_{i, 1}\rangle\|^2+\ldots+\||\psi_{i, t}\rangle\|^2 \geq
\frac{ (\||\psi_{i, 1}\rangle\|+\ldots+\||\psi_{i, t}\rangle\|)^2 }{t} .\]
Therefore,
\[ \||\psi_i\rangle\|\geq 
\frac{ \||\psi_{i, 1}\rangle\|+\ldots+\||\psi_{i, t}\rangle\|}{\sqrt{t}} .\]
$S$ is just a weighted sum of $\||\psi_i\rangle\|$ and
$S_1$, $\ldots$, $S_t$ are weighted sums of $\||\psi_{i, 1}\rangle\|$,
$\ldots$, $\||\psi_{i, t}\rangle\|$, respectively.
Hence, $S\geq \frac{S_1+\ldots+S_t}{\sqrt{t}}$ and 
\[ S_1+\ldots+S_t\leq \sqrt{t} S .\]
By definition of $r$, $S_r$ is the smallest of $S_1$, $\ldots$, $S_t$.
This implies (\ref{e5}).
$\Box$

Claim \ref{C3} gives 
\[ S'\leq S_r+\frac{2}{q-1}S \leq (\frac{1}{\sqrt{t}}+\frac{2}{q-1}) S = q' S.\]
This completes the proof of Lemma \ref{L1}.
$\Box$

Next, we use the $subdivide$ procedure to construct two inputs that are not 
distinguished by $A$. This is done as follows.
Let $v$ be $\lceil(\log(\frac{1}{10}(1- q(q')^u)
(1-\frac{1}{q}))/\log q'\rceil$.

\begin{enumerate}
\item[1.]
Let $m=n$, $l=1$, $s=1$.
\item[2.]
While $m\geq t^{v}$, repeat:
\begin{enumerate}
\item[2.1.]
$u$ times do $subdivide(m, l, s)$ and set $l=l'$, $m=m'$.
\item[2.2.]
$s=s+1$.
\end{enumerate}
\item[3.]
$v-u$ times do $subdivide(m, l, s)$ and set $l=l'$, $m=m'$.
\end{enumerate}

At the beginning, $m=n$. 
Each execution of step 2 decreases $m$ by a factor of $t^u$ and
step 2 is repeated while $m\geq t^v$.
Hence, it gets repeated at least 
\[ \frac{\log(n/t^v)}{\log (t^u)} =
\frac{\log n-v\log t}{u\log t}=\frac{\log n}{u\log t}-\frac{v}{u}=
\frac{\log n}{u\log t}-O(1) \]
times.

The final interval $[(l-1)m+1, lm]$
has a small probability of being queried and, therefore
it is impossible to distinguish $x_1=\ldots=x_k=0$,
$x_{k+1}=\ldots x_{n}=1$ for different $k\in [(l-1)m+1, lm]$
one from another. 
To prove this, we first show the following invariant:

\begin{Lemma}
\begin{enumerate}
\item
At the beginning of step 2.1., $S$ is at most $\frac{1}{1- q (q')^u}$.
\item
At the end of step 2.1., $S$ is at most $\frac{(q')^u}{1-q(q')^u}$.
\end{enumerate}
\end{Lemma}

\noindent
{\bf Proof:}
By induction. When we first start step 2.1., $s=1$, 
$S=\||\psi_1\rangle\|\leq 1$ and $1<\frac{1}{1- q(q')^u}$.

For the inductive case, if we have $S\leq\frac{1}{1- q (q')^u}$
at the beginning of step, each $subdivide$ decreases it $q'$ times
and $S\leq\frac{(q')^u}{1-q(q')^u}$ at the end of the step.
Also, if 
\[ S=\sum_{i=1}^s q^{s-i}\||\psi_i\rangle\| \]
at the end of one step, then, in the next step, $S$ will be
\[ \sum_{i=1}^{s+1} q^{s+1-i}\||\psi_i\rangle\|=
\||\psi_{s+1}\rangle\|+ q \sum_{i=1}^s q^{s-i}\||\psi_i\rangle\| 
\leq 1+ q \frac{(q')^u}{1-q (q')^u}=\frac{1}{1- q (q')^u}.\]
This completes the proof of the lemma.
$\Box$

By the same argument, $S\leq \frac{(q')^v}{1-q(q')^u}$ at the end of
step 3. The definition of $v$ implies
\[ S\leq \frac{\frac{1}{10}(1-q (q')^u)
(1-\frac{1}{q})}{1-q(q')^u}=\frac{1}{10}(1-\frac{1}{q}).\]
Together with $S=\sum_{i=1}^{s} q^{s-i} \| |\psi_i\rangle\| 
\geq q^{s-i} \| |\psi_i\rangle\| $, this implies
\begin{equation}
\label{e1} 
\||\psi_i\rangle\| \leq \frac{1-\frac{1}{q}}{10 q^{s-i}} .
\end{equation}

Now, we use a hybrid argument similar to \cite{BBBV,Sipser}.
Consider $A$ working on the input $x_1=\ldots=x_{lm}=0$,
$x_{lm+1}=\ldots=x_n=1$ and on the input $x_1=\ldots=x_{lm-1}=0$,
$x_{lm}=\ldots=x_n=1$. The final superpositions on these two inputs
are 
\[ |\varphi\rangle=U_T O_{lm} U_{T-1} \ldots U_0 (|0\rangle),
|\varphi'\rangle=U_T O_{lm-1} U_{T-1} \ldots U_0 (|0\rangle).\]
We are going to show that $\varphi$ and $\varphi'$ are close.
To show this, we introduce intermediate superpositions 
(also called hybrids)
\[ |\varphi_i\rangle =U_T O_{lm-1} U_{T-1} \ldots U_{i+1} O_{lm-1} U_i O_{lm} 
U_{i-1} \ldots U_0 (|0\rangle).\]
Then, $\varphi=\varphi_s$, $\varphi'=\varphi_0$. 

\begin{Claim}
\begin{equation}
\label{e2}
\||\varphi_{i}\rangle-|\varphi_{i-1}\rangle\|
\leq \frac{1-\frac{1}{q}}{5 q^{s-i}} .
\end{equation}
\end{Claim}

\noindent
{\bf Proof:}
The only different transformation is the $i^{\rm th}$ query
which is $O_{lm}$ for $|\varphi_i\rangle$ and $O_{lm-1}$ for 
$|\varphi_{i-1}\rangle$.
Before this transformation, the state is
\[ |\phi_i\rangle = U_{i-1} O_{lm} U_{i-2} \ldots U_0 (|0\rangle) \]
The part of this superposition corresponding to querying $[(l-1)m+1, lm]$
is $|\psi_i\rangle$. This is the only part of $|\phi_i\rangle$
on which $O_{lm}$ and $O_{lm-1}$ are different.
Therefore, 
\[ \| O_{lm}(|\phi_i\rangle)-O_{lm-1}(|\phi_i\rangle)\|=
 \| O_{lm}(|\psi_i\rangle)-O_{lm-1}(|\psi_i\rangle)\|\leq 
 2 \||\psi_i\rangle\| \leq 
 \frac{1}{5}\frac{1-\frac{1}{q}}{q^{s-i}} .\]
The next transformations ($U_T O_{lm-1} U_{T-1} \ldots U_{i}$)
are the same again. This implies (\ref{e2}).
$\Box$

By triangle inequality (and formulas (\ref{e2}) and (\ref{geom})),
\[ \||\varphi_{0}\rangle-|\varphi_s\rangle\| \leq 
\sum_{i=1}^{s} \||\varphi_{i-1}\rangle-|\varphi_{i}\rangle\| \leq
\sum_{i=0}^{s-1} \frac{1-\frac{1}{q}}{5 q^i} =
\frac{1-\frac{1}{q}}{5} \sum_{i=0}^{s-1}\frac{1}{q^{i}} \leq
\frac{1-\frac{1}{q}}{5}\frac{1}{1-\frac{1}{q}}=\frac{1}{5} .\]
However, $|\varphi_0\rangle$ is the final superposition for the input 
$x_1=\ldots=x_{lm-1}=0$, $x_{lm}=\ldots=x_n=1$, 
$|\varphi_s\rangle$ is the final superposition for the input 
$x_1=\ldots=x_{lm}=0$, $x_{lm+1}=\ldots=x_n=1$
and, by Lemma \ref{distance}, the distance between them must
be at least 1/4.

This shows that there is no quantum algorithm $A$ that solves 
the binary search problem with at most
$\frac{\log n}{u\log t}-\frac{v}{u}$ queries.
$\Box$

By optimizing the parameters in theorem \ref{main}, we get

\begin{Corollary}
At least $\frac{1}{12}\log n-O(1)$ queries are necessary for 
quantum binary search on $n$ elements.
\end{Corollary}

\noindent
{\bf Proof:}
Substitute $q=18.3$, $t=8$, $u=4$ into Theorem \ref{main}.
$\Box$

\section{Conclusion}

We have shown that any quantum algorithm needs at least $\log_2 n/12$ 
queries to do binary search. This shows that at most a constant speedup
is possible for this problem in the query model
(compared to the best classical algorithm).

Similarly to other lower bounds on quantum algorithms, this
result should not be considered as pessimistic. 
First, the classical binary search is very sequential algorithm and,
therefore, it is not so surprising that it is impossible to
speed it up by using quantum algorithms.
Second, the classical binary search is already fast enough for
most (if not all) practical purposes.


We hope that our lower bound technique will be useful for 
proving other lower bounds on quantum algorithms. 
One of main open problems in this area is the 
collision problem\cite{Collision} 
for which there is no quantum lower bounds at all.

{\bf Acknowledgments.}
I thank Ashwin Nayak for suggesting this problem and Ronald
de Wolf for useful comments.

\end{document}